\documentstyle[twocolumn,floats,epsfig,prb,aps]{revtex}
\begin{document}
\draft 
\title{Optical transitions in broken gap heterostructures}
\author{E. Halvorsen\cite{EHa}}
\address{Department of Physics, P.O. Box 1048 Blindern, N-0316 Oslo,
Norway} 
\author{Y. Galperin}
\address{Department of Physics, P.O. Box 1048 Blindern, N-0316 Oslo,
Norway,
\\ and Division of Solid State Physics, Ioffe Physico-Technical Institute
of Russian Academy of Sciences, \\ 194021 St. Petersburg, Russia}
\author{K. A. Chao}
\address{Department of Physics, Lund University, S\"olvegatan 14A,
S-223 62 Lund, Sweden}

\date{\today}
\maketitle
\begin{abstract}
We have used an eight band model to investigate the electronic structures
and to calculate the optical matrix elements of InAs-GaSb broken gap
semiconductor heterostructures. The unusual hybridization of the conduction
band states in InAs layers with the valence band states in GaSb layers has
been analyzed in details. We have studied the dependence of optical matrix
elements on the degree of conduction-valence hybridization, the tuning of
hybridization by varying the width of the GaSb layers and/or InAs layers,
and the sensitivity of quantized levels to this tuning. Large spin-orbit
splitting in energy bands has been demonstrated. Our calculation can serve
as a theoretical modeling for infrared lasers based on broken gap quantum
well heterostructures.
\end{abstract}
 
\section{Introduction} 

The characteristic feature of Al$_x$Ga$_{1-x}$Sb/InAs heterostructures with
$x$$<$0.3 is the overlap of the InAs conduction band with the
Al$_x$Ga$_{1-x}$Sb valence band~\cite{a1}. Such systems, often referred to
as {\it broken gap} heterostructures, exhibit interesting negative
persistent photoconductivity~\cite{a2}, semimetal-semiconductor transition
induced by magnetic field~\cite{a3}, and intrinsic exciton~\cite{a4}. As the
Al concentration increases from $x$=0.3, a staggered band alignment appears
at the Al$_x$Ga$_{1-x}$Sb/InAs interface with the valence band edge of
Al$_x$Ga$_{1-x}$Sb lying in the band gap of InAs. On the other hand, at the
GaSb/Al$_x$Ga$_{1-x}$Sb interface the band alignment is of the straddled
type. Therefore, in a InAs/Al$_x$Ga$_{1-x}$Sb/GaSb heterostructure, carriers
can tunnel from the InAs conduction band to the GaSb valence through the
Al$_x$Ga$_{1-x}$Sb barrier.

InAs, GaSb, and AlSb form a family of semiconductors with sufficient
lattice match for epitaxy growth. Interband tunneling devices based on the
InAs/AlGaSb/GaSb double barrier resonant tunneling have been fabricated,
which exhibit high frequency response and peak-to-valley current ratio at
room temperatures~\cite{a4,a5,a6,a7,a8,a9,a10,a11,a12}. Because of their
potential for technological application, resonant
tunneling~\cite{a13,a14,a15} and resonant
magnetotunneling~\cite{a16,a17,a18,a19} have been investigated extensively.

\begin{figure}[ht]
\begin{center}
\epsfig{file=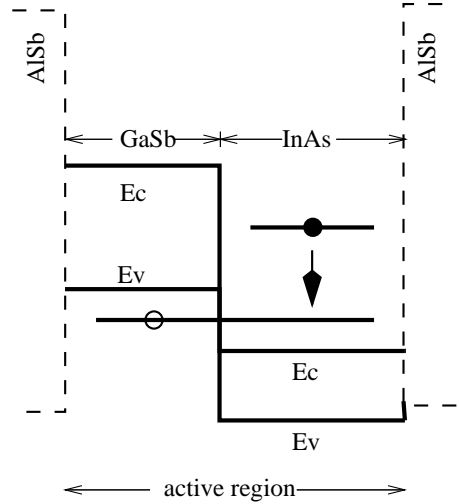,width=6cm}
\end{center}
\caption{The possible hybridization between the InAs conduction band
states and the GaSb valence band states in the active region of a broken
gap quantum well structure provides the novel infrared radiation.}
\label{fig_opprinc}
\end{figure}

To the contrary of tunneling processes, very little work has been done on
the electronic structure and optical properties of broken gap
heterostructures~\cite{ChangYC1985,a20,a21}. We notice that due to the
overlap of InAs conduction band and the GaSb valence band, it is possible
to form new type of eigenstates with interesting optical properties. We
will elaborate this with Fig.~\ref{fig_opprinc} where the conduction band
edge $E_c$ and the valence band edge $E_v$ are marked for a GaSb/InAs
heterostructure embedded in the band gap of AlSb. Treating the AlSb at both
sides as potential barriers, the system AlSb/GaSb/InAs/AlSb is essentially
a new type of quantum well which we name as broken gap quantum well (BGQW).
If we diminish the thickness of either the GaSb layer or the InAs layer to
zero, the BGQW reduces to a conventional quantum well which, with bipolar
doping becomes a quantum well laser.

If in a properly designed BGQW there exist two eigenenergies as shown in
Fig.~\ref{fig_opprinc}, under bipolar doping, the BGQW can lase in the
infrared frequency region. Such semiconductor infrared lasers or detectors
were proposed earlier~\cite{YangRQ1991,OhnoH1992}. The lower energy state
in Fig.~\ref{fig_opprinc} is characterized by the degree of hybridization
between the GaSb valence band states and the InAs conduction band states.
The enhancement of optical transition requires the hybridized wave function
having a large amplitude in the InAs layer. On the other hand, to drain out
the electrons in the hybridized state in order to acheive the population
inversion for lasing, the wave function should have a large component from
the GaSb layer. A thorough understanding of this hybridization is important
not only for opto-electronic devices, but also for the fundamental theory
of transport parallel to interfaces.

The optical property of a BGQW, which is of most theoretical interest and
of most importance to opto-electronics, involves the optical transition
between these two levels. Hence, relevant issues to be studied are the
dependence of optical matrix elements on the degree of conduction-valence
hybridization, the tuning of hybridization by varying the width of the
GaSb layer, and the sensitivity of quantized levels to this tuning. In
this paper we will perform a theoretical calculation of electronic
structure and optical matrix elements in order to investigate these items.
In Sec.~\ref{sect_model} we present the model system we study and the
basic assumptions in our theoretical calculation. Our results for the
electronic structure and optical matrix elements are  given in
Sec.~\ref{sect_elstruct} and Sec.~\ref{sect_optmat} respectively, and
discussed in Sec.~\ref{sect_discuss}.

\section{Model and theory} \label{sect_model}

Our model system shown in Fig.~\ref{fig_opprinc} is a GaSb/InAs
heterostructure sandwiched between two thick AlSb layers. The growth
direction is [001] which defines the $z$ axis. The $x$ axis is along
[100] and the $y$ axis is along [010]. This system was studied earlier
with a simple two band model~\cite{NavehY1995}. The essential feature of
our model is the existence in BGQW quantized conduction band states in
the InAs layer, and hybridized quantum levels in the energy regime where
the GaSb valence band overlaps with the InAs conduction band. Therefore,
for simplicity we can well assume that the two thick AlSb layers are bulk
AlSb. As an illustration, one conduction band level and one hybridized
level are plotted in Fig.~\ref{fig_opprinc}. We will perform band
structure calculation for the BGQW shown in Fig.~\ref{fig_opprinc}. The
lattice constants are 6.096 {\AA} for GaSb, 6.058 {\AA} for InAs, and
6.136 {\AA} for AlSb. With such small lattice mismatch, the strain
estimated with the deformation potential of these materials is not large
enough to modify our conclusion qualitatively. Hence, in our theoretical
calculation, strain will be neglected. As will be discussed in
Sec.~\ref{sect_elstruct}, the essential feature of our model mentioned
above, and hence the conclusion reached by our calculation will remain
intact even this weak strain effect is taken into account.

We employ the effective bond orbital model~\cite{ChangYC1988} (EBOM) which
is a tight-binding-like model defined on the Bravais lattice of the
underlying crystal. The basis are constructed as linear combinations of
orbitals centered on each lattice site having $s$- or $p$-symmetry, together
with spin eigenstates. The so-constructed basis functions diagonalize the
spin-orbit interaction. Our model Hamiltonian includes local and nearest
neighbor matrix elements. The degree of inversion asymmetry of the BGQW
heterostructure under study is much higher than that of individual InAs
or GaSb constituent layers. Hence, in our calculation the effect of weak
inversion symmetry is neglected for simplicity. On the other hand, the
strong inversion symmetry of the BGQW and its combined effect with the
spin-orbit coupling at finite in-plane wavevector are fully taken into
account in our theoretical treatment.

It is worthwhile to point out that the EBOM is not an atomistic model like
the tight binding model. While the tight binding model is built from
orbitals representing the degree of freedom of each individual atom, the
bond-orbitals in the EBOM are confined to a primitive unit cell and
represent more than one atom. The symmetry of the EBOM is identical to
that of the {\bf k}$\cdot${\bf p} model, and therefore in both models there
is no asymmetry between the (110) and (1$\bar 1$0) directions. On the other
hand, the difference between the EBOM and the {\bf k}$\cdot${\bf p} model
is that the {\bf k}$\cdot${\bf p} model is a continum model without a
natural momentum cutoff, but the EBOM is discretized on the Bravais lattice
of the crystal.

The bulk parameters of the EBOM, which are required for our calculation,
are obtained by fitting the bulk band-structure of EBOM to that of the
{\bf k}$\cdot${\bf p} Hamiltonian near the $\Gamma$ point to the second
order in $k$, as described in Ref~\onlinecite{ChangYC1988}. The EBOM can
therefore be considered as an effective mass theory discretized on the
Bravais lattice. For small {\bf k} vectors the two models are equivalent.
However, since EBOM is discretized on the Bravais lattice, it gives better
results at large $k$. It is important to point out that usually, as in the
present case, it is not even necessary to know the precise form of the
effective bond orbitals.

Because of the large overlap between the InAs conduction band and the GaSb
valence band, the split-off band must be included in our treatment. Hence,
we will deal essentially an eight band Kane model. We consider the BGQW
heterostructures grown in the [001] direction, with $n$ labelling the
(001) planes. Then, the position of a lattice site $\vec R_n$ in the $n$th
plane can be expressed as $(n\frac{a}{2},\vec R_n)$, where $a$ is the
lattice constant of the conventional unit cell. The eight bond orbitals at
each lattice site are conventionally labelled as
($S\frac{1}{2}\frac{1}{2}$),
($S\frac{1}{2}-\frac{1}{2}$),
($P\frac{3}{2}\frac{3}{2}$),
($P\frac{3}{2}\frac{1}{2}$),
($P\frac{3}{2}-\frac{1}{2}$),
($P\frac{3}{2}-\frac{3}{2}$),
($P\frac{1}{2}\frac{1}{2}$), and
($P\frac{1}{2}-\frac{1}{2}$).
We denote $|n,\vec R_n,\alpha\rangle$ for the $\alpha$th bond orbital at
position $(n\frac{a}{2},\vec R_n)$. In terms of these bond orbitals, we
define a basis set of planar orbitals
\begin{equation}
|n,\vec k_\parallel,\alpha \rangle = 
	N_{\parallel}^{-1/2}\sum_{\vec R_n}
        \exp(i\vec k_\parallel\cdot\vec R_n)
        | n,\vec R_n,\alpha  \rangle \,,
\end{equation}
where $\vec k_\parallel$ is an in-plane wave vector, and $N_{\parallel}$
is the number of sites in each plane.

Using this basis of planar orbitals, the Hamiltonian
\[
H = \sum_{\vec k_\parallel} H_{\vec k_\parallel}
\]
is decomposed into a linear combination of partial Hamiltonians
$H_{\vec k_\parallel}$. The partial Hamiltonian
\begin{eqnarray}\label{ph}
H_{\vec k_\parallel} & = &  
\sum_{n \alpha \beta} e_{n,\vec k_\parallel,\alpha\beta}
|n,\vec k_\parallel,\alpha\rangle\langle n,\vec k_\parallel,\beta|
\nonumber \\
& & +  \sum_{n \alpha \beta}
\left( v_{n,\vec k_\parallel,\alpha\beta}
|n+1,\vec k_\parallel,\alpha\rangle\langle n,\vec k_\parallel,\beta|
+ h.c. \right)
\end{eqnarray}
is diagonal with respect to the in-plane wave vectors $\vec k_\parallel$,
and block-tridiagonal with respect to the remaining quantum numbers. This
form of Hamiltonian is suitable for numerical computations.

The matrix elements in $H_{\vec k_\parallel}$, except for those
$v_{n,\vec k_\parallel,\alpha\beta}$ which connect the two bond orbital
planes forming an interface, are set to the values for the corresponding
bulk  materials. At an interface, we follow the commonly accepted approach
to determine the value of $v_{n,\vec k_\parallel,\alpha\beta}$ as the
average of the parameter values of the bulk materials on each side of the
interface. The materials parameter values from which the effective bond
orbital parameters are derived are given in Table~\ref{tab_kppar}. The
valence band offsets are 0.56 eV for GaSb-InAs, 0.18 eV for AlSb-InAs,
and -0.38 eV for AlSb-GaSb.
\begin{table}[ht]
\caption{Parameters used to determine effective bond orbital parameters}
\label{tab_kppar}
\begin{tabular}{lddd}
 & InAs & GaSb & AlSb  \\ \hline
 $a$ (\AA)  &  6.0583 & 6.082 & 6.133 \\
 $E_g$ (eV) & 0.41 & 0.8128 & 2.32 \\
 $m_c/m_0$ & 0.024 & 0.042 & 0.18  \\
 $\Delta$ (eV) & 0.38 & 0.752 & 0.75 \\
 $E_P$ (eV) & 22.2 & 22.4 & 18.7 \\
 $ \gamma_1 $ & 19.67 & 11.80 & 4.15 \\
 $ \gamma_2 $ & 8.37 & 4.03 & 1.01 \\
 $ \gamma_3 $ & 9.29 & 5.26 & 1.75 
\end{tabular}
\end{table}

For a given $\vec k_\parallel$, the eight eigenfunctions 
\[
|\Psi_{\gamma,\vec k_\parallel}\rangle =
\sum_{n,\alpha} F^{\gamma}_{n,\vec k_\parallel,\alpha}
|n,\vec k_\parallel,\alpha\rangle
\,; \,\,\, \gamma=1,2,\cdots
\]
and the corresponding band energies $E_{\gamma,\vec k_\parallel}$ can be
readily obtained from Eq.~(\ref{ph}) by diagonalizing a finite tridiagonal
matrix for a BGQW of finite width. Knowing the eigensolutions, we will
calculate the optical matrix elements.

The representation of the momentum operator in the bond-orbital basis 
contains the leading local and nearest neighbor matrix elements. They are
determined by mapping the bulk matrix elements to the
{\bf k}$\cdot${\bf p} results up to and including terms linear in wave
vector~\cite{ChangYC1990}. In the planar orbital basis, its
$\vec k_\parallel$-diagonal part can be expressed as
\begin{eqnarray}\label{ome}
P^\nu(\vec k_\parallel) & = &  
\sum_{n \alpha \beta} P^\nu_{n,\vec k_\parallel,\alpha\beta}
|n,\vec k_\parallel,\alpha\rangle\langle n,\vec k_\parallel,\beta| \\
& & + \sum_{n \alpha \beta} \left( Q^\nu_{n,\vec k_\parallel,\alpha\beta}
|n+1,\vec k_\parallel,\alpha\rangle\langle n,\vec k_\parallel,\beta|
+ h.c. \right) \,, \nonumber
\end{eqnarray}
where $\nu$=$x,y,z$ is the polarization direction. The matrix elements
are
\begin{eqnarray}
P^\nu_{n,\vec k_\parallel,\alpha\beta} & = &
\sqrt{\frac{2}{m_0}} \langle n,\vec k_\parallel,\alpha |p_\nu 
|n,\vec k_\parallel,\beta \rangle \, , \\
Q^\nu_{n,\vec k_\parallel,\alpha\beta} & = &
\sqrt{\frac{2}{m_0}} \langle n+1,\vec k_\parallel,\alpha |p_\nu
| n,\vec k_\parallel,\beta \rangle \, ,
\end{eqnarray}
where $m_0$ is the free electron mass. In the above matrix elements we
have included the prefactor $\sqrt{2/m_0}$ such that these matrix elements
have units $(eV)^{1/2}$.

\section{Electronic structure} \label{sect_elstruct}

In the following we will examine the electronic structure of the InAs-GaSb
BGQW shown in Fig.~\ref{fig_opprinc}. Before the InAs layers and the GaSb
layers are coupled together, we will denote the $n$th electronic levels in
the conduction band by En, the $n$th energy levels in the heavy-hole band
by Hn, and the $n$th energy levels in the light-hole band by Ln. Such
conventional labeling is unambiguous for $\vec k_\parallel$=0.

\begin{figure}[ht]
\begin{center}
\epsfig{file=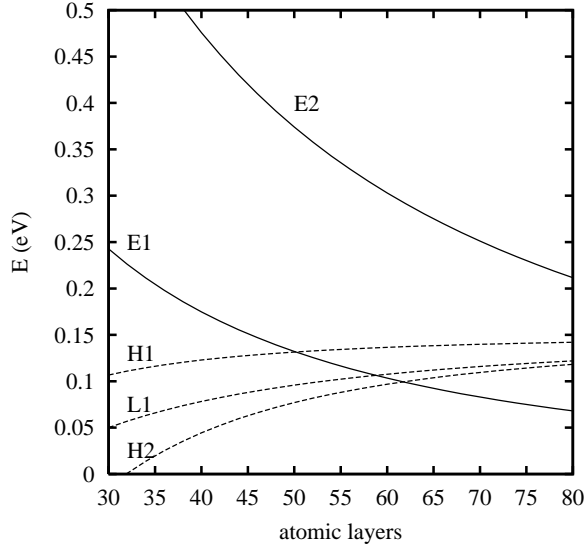,width=11cm}
\end{center}
\caption{Electron energies in a AlSb-InAs-AlSb quantum well (solid curves),
and in a AlSb-GaSb-AlSb quantum well (dashed curves) as functions of well
width. All energies are measured from the InAs conduction band edge.}
\label{fig_swlev}
\end{figure}

The symmetry properties of EBOM at $\vec k_\parallel$=0 allows the
conduction band states in the InAs layers to hybridize with the light-hole
band states in the GaSb layers, but not with the heavy-hole band states. We
would like first to locate the region of BGQW structures in which we can
fine tune the degree of such hybridization. For this purpose, we perform
two energy level calculations; one with zero InAs layer width and the other
with zero GaSb layer width. The results are plotted in Fig.~\ref{fig_swlev}
as functions of the number of atomic layers. We see that for 58 atomic
layers, the GaSb L1 level overlaps with the InAs E1 level. Hence, the
region around 60 atomic layers will be the reasonable starting BGQW
structure for tuning the system into maximum hybridization. We should be
aware of the fact that different confinement boundary conditions also
affect the degree of hybridization, as will be seen in our computed results
below.

\begin{figure}[ht]
\begin{center}
\epsfig{file=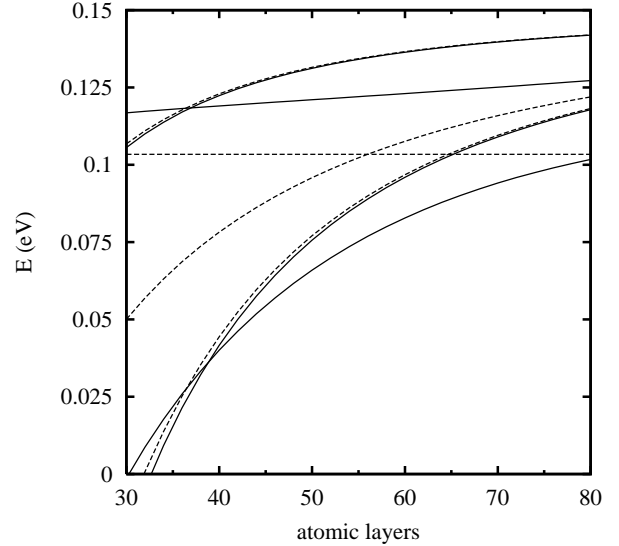,width=11cm}
\end{center}
\caption{Zone center ($\vec k_\parallel$=0) H1, H2, L1, and E1 levels as
functions of GaSb layer width in a InAs-GaSb BGQW with 60 InAs atomic
layers. Dashed curves are for the case without the InAs-GaSb interface
coupling (ordered as H1$>$L1$>$H2$>$E1 at 80 atomic layers of GaSb), and
the solid curves are for the case with the InAs-GaSb interface coupling
(for convenience, ordered as H1$>$E1$>$H2$>$L1 at 80 atomic layers of
GaSb).}
\label{fig_e0i60}
\end{figure}

We will then set 60 atomic layers for the InAs constituent and vary the
GaSb width from 30 to 80 atomic layers. In the absence of interface
coupling, the H1, E1, H2 and L1 levels at $\vec k_\parallel$=0 are shown in
Fig.~\ref{fig_e0i60} as dashed curves (hole states are the same as in
Fig.~\ref{fig_swlev}). In the BGQW sample with 80 atomic layers of GaSb,
these four levels are ordered as H1$>$L1$>$H2$>$E1. When the interface
coupling is turned, these four levels are shown in Fig.~\ref{fig_e0i60} as
solid curves. For the convenience of description, we use the results at 80
atomic layers of GaSb to label them as H1$>$E1$>$H2$>$L1. Because of the
symmetry properties at $\vec k_\parallel$=0, the conduction band states
have negligible influence on H1 and H2 levels. The heavy hole levels are
only slightly perturbed by the change of boundary conditions. On the other
hand, the L1 and E1 levels are strongly hybridized and repel each other.
The difference between the solid curves and the dashed curves is also due
to the different confinement boundary conditions as mentioned above. In the
region of GaSb width between 50 and 60 atomic layers, the separation of the
two resulting hybridized levels is about 40-55 meV.

\begin{figure}[ht]
\begin{center}
\epsfig{file=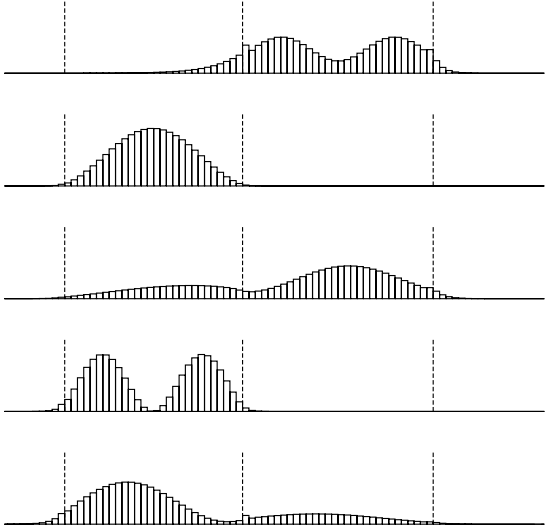,width=7.5cm}
\end{center}
\caption{Real space occupation probability of five zone center
($\vec k_\parallel$=0) eigenstates in the active region of a BGQW with
56 GaSb atomic layers and 60 InAs atomic layers.}
\label{fig_wf0g56i60}
\end{figure}

\begin{figure}[ht]
\begin{center}
\epsfig{file=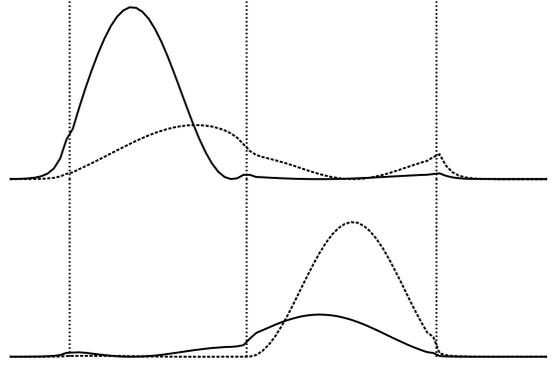,angle=270,width=7.5cm}
\end{center}
\caption{Structure of the states L1 (solid curves) and E1 (dotted curves),
with the two upper plots for the $P_{\frac{3}{2},\pm\frac{1}{2}}$
components and the two lower plots for the $S_{\frac{1}{2},\pm\frac{1}{2}}$
components.
\label{fig_Tmpfig}}
\end{figure}

To demonstrate the spatial properties of hybridized states, we have
calculated the real space occupation probability
\[
{\cal O}^{\gamma}_n(\vec k_\parallel=0) \equiv
\sum_{\alpha} |F^{\gamma}_{n,\vec k_\parallel=0,\alpha}|^2
\,; \,\,\, \gamma=1,2,\cdots
\]
${\cal O}^{\gamma}_n(\vec k_\parallel$=0) of five eigenstates in a BGQW
with 56 GaSb atomic layers and 60 InAs atomic layers are plotted in
Fig.~\ref{fig_wf0g56i60}. From top to bottom, the first state is E2 which
is largely confined in InAs layers. The second and fourth state, which
localized in GaSb layers, are H1 and H2 respectively. The third and the
lowest states are E1-L1 hybridized states. To be consistent with our
earlier convention, the third state is E1, and the lowest state is L1. One
way to further analyze this issue is to decompose the total occupation
probability into partial occupation probability for each $\alpha$ band.
Such decomposition is shown in Fig.~\ref{fig_Tmpfig} with dotted curves
for the E1 state, and the solid curves for the L1 state. In this figure,
the two upper plots are for the $P_{\frac{3}{2},\pm\frac{1}{2}}$
components and the two lower plots are for the
$S_{\frac{1}{2},\pm\frac{1}{2}}$ components. We see that the tail of the
E1 (or L1) state in the GaSb (or InAs) region has the same component
profile as its main part in the InAs (or GaAs) region. Such features were
also obtained for InAs-GaSb superlattices in a very recent
paper~\cite{a21}.

In our derivation of the eigensolutions of a BGQW heterostructure, the
effect of strain has been neglected. The separations between the
so-obtained eigenenergies for $\vec k_\parallel$=0, which are determined
by the confinement potential, are substantially larger than the energy
shifts produced by the strain. Hence, even the effect of strain is
included in an improved calculation, the formation of E1-L1 hybridized
states is still dominated by the confinement potential, and the degree of
hybridization can still be tuned by changing the width of the GaSb and/or
the InAs layers. In this respect, the conclusion reached with our present
calculation will remain intact when the effect of strain is taken into
account.

\begin{figure}[ht]
\begin{center}
\epsfig{file=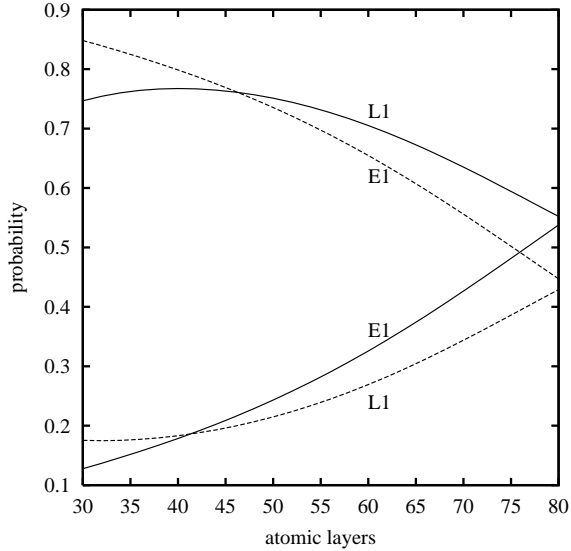,width=11cm}
\end{center}
\caption{ Probabilities of finding an electron in the GaSb layers (solid
lines) or in the InAs layers (dashed lines) as function of GaSB layer
width, when the electron occupies the hybridized states E1 or L1 in a
BGQW with 60 InAs atomic layers.}
\label{fig_wfani60}
\end{figure}

An interesting question is, when a charge carrier occupies a E1-L1
hybridized state, will its physical properties electron-like or hole-like?
For example, what will be the cyclotron effective mass of such a charge
carrier? One relevant quantity is the probability
\[
{\cal O}^{\gamma}_{\rm InAs} (\vec k_\parallel=0) \equiv
\sum_{n\in{\rm InAs}} {\cal O}^{\gamma}_n(\vec k_\parallel=0)
\,; \,\,\, \gamma=1,2,\cdots
\]
to find the electron in InAs layers when it occupies the $\gamma$th
eigenstate. For the two hybridized states (labeled as E1 and L1 in our
convention) in a InAs-GaSb BGQW with 60 InAs atomic layers, the results
are plotted in Fig.~\ref{fig_wfani60} as function of the number of GaSb
atomic layers. The solid curves are for
${\cal O}^{\gamma}_{\rm GaSb}(\vec k_\parallel$=0), and the dashed curves
are for ${\cal O}^{\gamma}_{\rm InAs}(\vec k_\parallel$=0). For thin GaSb
layers, the charge carrier in a hybridized state is mostly either
electron-like or hole-like. However, as the width of GaSb layers increases
towards 80 atomic layers, the characteristic features of the charge
carrier remain to be studied.

\begin{figure}[ht]
\begin{center}
\epsfig{file=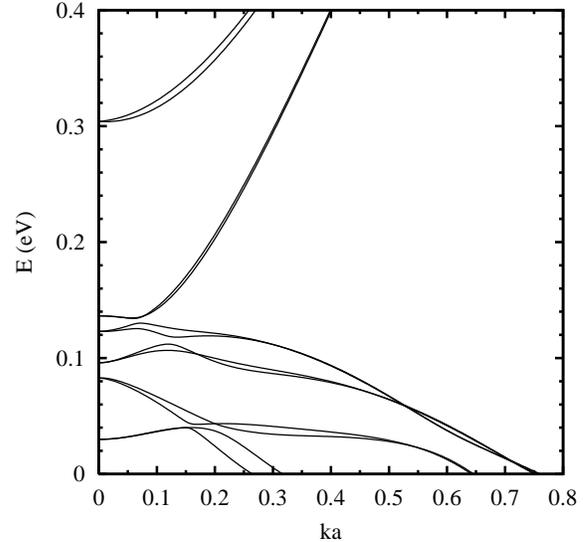,width=11cm}
\end{center}
\caption{Subband dispersion with $\vec k_\parallel$ along [100] direction
for an InAs-GaSb BGQW with 60 InAs atomic layers and 60 GaSb atomic
layers.}
\label{fig_rdg60i60}
\end{figure}

The above calculations for zone center ($\vec k_\parallel$=0) states are
repeated for finite $\vec k_\parallel$, and the dispersion relations of
various subbands are shown in Fig.~\ref{fig_rdg60i60} for a BGQW with
60 InAs atomic layers and 60 GaSb atomic layers. We have chosen
$\vec k_\parallel$ along the [100] direction, or the $x$ axis. At zone
center, from top to bottom the levels are E2, H1, E1, H2, L1, and L2. In
region of finite $\vec k_\parallel$, the spin-splitting of levels and the
anticrossing of levels are clearly seen. We will return to these
dispersion relations later for further discussions.

We have mentioned that our BGQW system was studied earlier with a simple
two-band model~\cite{NavehY1995}, where only the H1 and the E1 levels are
included. Therefore, in Ref.~\onlinecite{NavehY1995} there is no E1-L1
hybridization. The energy bands shown in Fig.~2 of
Ref.~\onlinecite{NavehY1995} corresponds to our E1 and H1 bands in
Fig.~\ref{fig_rdg60i60}, within the region 0$<$$ka$$<$0.12. The two-band
calculation does yield an anticrossing at finite $\vec k_\parallel$, but
gives no spin-orbit splitting.

\section{Optical matrix elements} \label{sect_optmat}

Knowing the eigenfunctions of the BGQW heterostructure, the optical matrix
elements are readily calculated with Eq.~(\ref{ome}). Because for finite
$\vec k_\parallel$ each dispersion curve is spin-orbit splited into two
curves, we distinguish them as Ena and Enb for conduction band states, Hna
and Hnb for heavy-hole band states, and Lna and Lnb for light-hole band
states. For convenience, in the region of small $\vec k_\parallel$, the
lower spin-orbit splited dispersion curve is assigned with a, and the
higher one with b.

There are numerous optical transitions between each pair of states for
different polarizations of the electromagnetic waves. The selection rules
are complicated by the anticrossing of levels as well as the
hybridization between InAs conduction band states and GaSb valence band
states. Hence, here our study will focus on the cases of our interest: the
optical transitions from the E2a and E2b levels to the E1a, E1b, H1a, and
H1b levels. As will be discussed later, such transitions are relevant to
the possible infrared lasers based on BGQW heterostructures. We would like
to mention that parallel to our work, in a very recent paper~\cite{a21},
in InAs-GaSb superlattices the optical transition matrix elements at
$\vec k_\parallel$=0 with in-plane polarization were studied as functions
of the superlattice period.

By analyzing our extensive numerical results, we have reached the
following selection rules for transitions from the E2a and E2b levels to
the E1a, E1b, H1a, and H1b levels. For $\vec k_\parallel$ in the [100] or
the [110] direction, among the two transitions from E2a (or E2b) to E1a or
E1b, only one is allowed. Similarly, among the two transitions from E2a
(or E2b) to H1a or H1b, only one is allowed. However, if
$\vec k_\parallel$ is along a low symmetry direction in the
two-dimensional Brillouin zone, such as [210] for example, all transitions
are allowed. It is worthwhile to point out that these selection rules for
a BGQW, which is asymmetric, happen to be the same as the selection rules
for the intersubband transitions in a symmetric conduction band quantum
well, derived with a simplified eight band model~\cite{YangRQ1994}.

\begin{figure}[ht]
\begin{center}
\epsfig{file=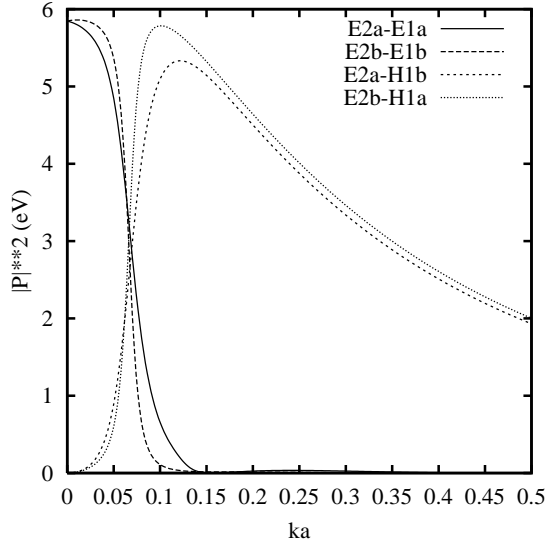,width=11cm}
\end{center}
\caption{Square of the amplitude of optical matrix elements with
$z$-polarization as functions of $\vec k_\parallel$ which is along $x$
axis, for a BGQW with 60 InAs atomic layers and 60 GaSb atomic layers.}
\label{fig_optmatz60i60}
\end{figure}

The numerical results to be discussed below are obtained for a BGQW with
60 InAs atomic layers and 60 GaSb atomic layers, the band structure of
which is given in Fig.~\ref{fig_rdg60i60}. From Fig.~\ref{fig_wfani60}
we see a significant E1-L1 hybridization in this BGQW heterostructure.
The square of the amplitude of optical matrix elements with
$z$-polarization are shown in Fig.~\ref{fig_optmatz60i60}. At
$\vec k_\parallel$=0, the optical matrix element is zero for the
E2$\rightarrow$H1 transition, but is large for the E2$\rightarrow$E1
transition, as expected from the symmetry properties. As
$\vec k_\parallel$ increases, the anticrossing between the E1 band and the
H1 band occurs around $\vec k_\parallel a$$\simeq$0.75. Consequently, the
E2$\rightarrow$E1 transition drops sharply to zero, while the
E2$\rightarrow$H1 transition picks up its strength rapidly.

\begin{figure}[ht]
\begin{center}
\epsfig{file=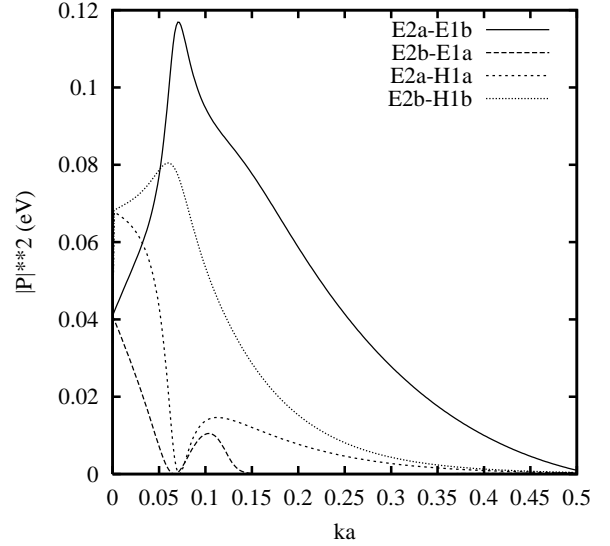,width=11cm}
\end{center}
\caption{Same as Fig.~\ref{fig_optmatz60i60} but with $y$-polarization
of the electromagnetic wave.}
\label{fig_optmaty60i60}
\end{figure}

\begin{figure}[ht]
\begin{center}
\epsfig{file=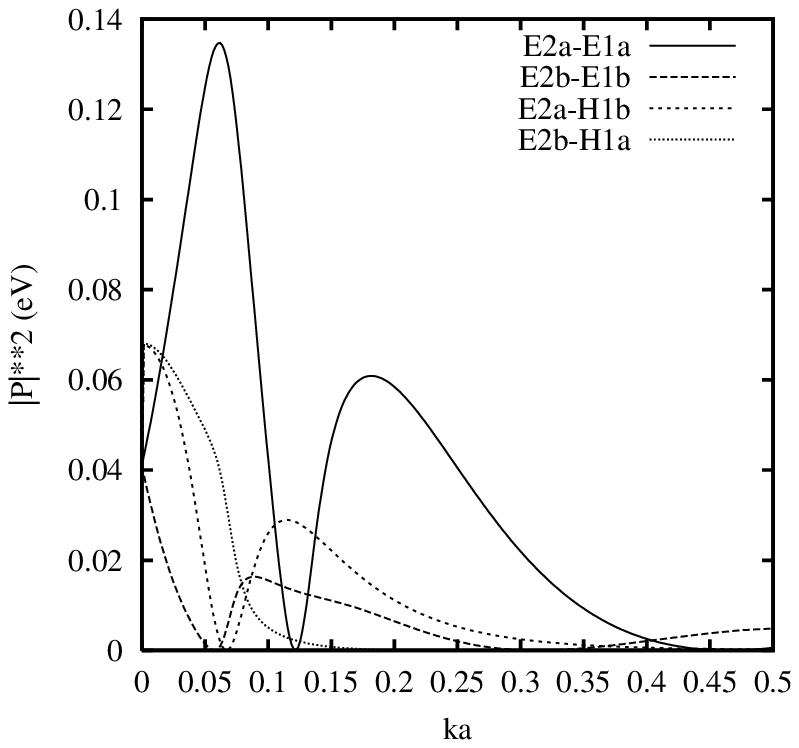,width=11cm}
\end{center}
\caption{Same as Fig.~\ref{fig_optmatz60i60} but with $x$-polarization
of the electromagnetic wave.}
\label{fig_optmatx60i60}
\end{figure}

By changing the polarization of the electromagnetic wave and repeating the
calculations of optical matrix elements, the results for $y$-polarization
is shown in Fig.~\ref{fig_optmaty60i60} and for $x$-polarization in
Fig.~\ref{fig_optmatx60i60}. The strength of these transitions are
substantially smaller than those for the $z$ polarization, by about two
orders of magnitude. The strong wavevector dependence of the optical
matrix elements in Figs.~\ref{fig_optmaty60i60} and \ref{fig_optmatx60i60}
is due to the combined effect of spin-orbit splitting and anticrossing of
energy bands. The drastic difference between the curves in
Fig.~\ref{fig_optmaty60i60} and the corresponding curves in
Fig.~\ref{fig_optmatx60i60} indicates the strong anisotropy of the optical
matrix elements with respect to in-plane polarizations.

\section{Discussion} \label{sect_discuss}

The very unusual feature of the BGQW heterostructures is the new
eigenstates formed by the hybridization of conduction band states at one
side of the interface with the valence band states at the other side. We
made a thorough investigation on the formation of such states and their
influence on the optical properties of a BGQW. However, their influence on
the transport properties parallel to interfaces is perhaps a more important
issue for fundamental study. If one tunes the the system to have the Fermi
energy lying in a conduction-valence hybridized two-dimensional energy
band, depending on the degree of hybridization, the parallel transport
properties can change from completely electron-like to complete hole-like.

Another relevant issue for fundamental study is the spin-orbit splitting.
Recently, there has been much interest in the effect of spin-orbit
interactions in two-dimensional electron gases. For example, the
Shubnikov-de Haas measurements of spin-orbit splitting have been performed
on symmetric InAs/GaSb~\cite{LuoJ1990} quantum wells and on symmetric
InAs/AlSb~\cite{HeidaJP1998} quantum wells, in which the origin of
spin-orbit interaction is the lack of inversion symmetry in bulk crystal
structure. Because of the asymmetry of the BGQW itself, very large
spin-orbit splittings appear in the energy bands as shown in
Fig.~\ref{fig_rdg60i60}. Consequently, BGQW heterostructures are good
candidate for investigating phenomena which are related to spin-orbit
interaction.

BGQW heterostructures have great potential in technological applications
for lasers and detectors, tunnable in infrared wavelengths~\cite{a23,a24}.
Our calculations can serve as the theoretical modeling of such infrared
lasers. To make a bipolar laser similar to the conventional quantum
lasers, in Fig.~\ref{fig_opprinc} the AlSb layers at the InAs side should
be $n$-doped, and the AlSb layers at the GaSb side should be $p$-doped.
The number of InAs atomic layers should be around 60, and the number of
GaSb atomic layers should be between 70 and 80. Then, from
Fig.~\ref{fig_wfani60} we see a strong E1-L1 hybridyzation. By adjusting
the acceptor concentration, we can set the Fermi level between the E1
level and the H2 level shwon in Fig.~\ref{fig_wf0g56i60}. When such BGQW
heterostructure is connected to external circuit, electrons which are
injected into the E2 energy band will relax to the zone center and make
radiation transitions to the E1 band due to the large optical matrix
elements of $z$-polarized light. Then, the strong conduction-valence
hybridization of the E1 band states provides the rapid draining of
electrons from the E1 band into external circuit via the GaSb valence band
states. From Fig.~\ref{fig_rdg60i60} we estimate the energy of emitted
photons to be 0.18-0.2 eV, corresponding to wavelength about 6 micrometers.
Around this wavelength there exist infrared windows, and consequently such
radiation source will be extremely useful.

To close this paper, we must mention the very recent work of
Ref.~\onlinecite{a21} in which the electronic structure and optical
matrix elements at the zone center ($\vec k_\parallel$=0) of InAs-GaSb
superlattices have been calculated with both a plane wave pseudopotential
method and the eight band {\bf k}$\cdot${\bf p} method. In that paper the
effects of superlattice period on various physical quantities at zone
center were studied in details. In particular, the authors of
Ref.~\onlinecite{a21} found a zone center E1-H1 coupling manifested
by band anticrossing at superlattice period $n$=28, as well as a zone
center L1-H2 coupling and anticrossing around $n$=13. Such features
are absent in the {\bf k}$\cdot${\bf p} method because it fails to
recognize the atomistic details in no-common-atom superlattice. How these
features will affect quantitatively the 2D dispersion relation, and hence
our results of the formation of E1-L1 hybridization and the in-plane
physical properties of a single BGQW heterostructure, as well as their
impact on the theoretical modeling of infrared lasers and detectors,
remains to be an open question.

\acknowledgments
This work was supported by the Norwegian Research Council (NFR) under 
grant no. 111071/431.


\begin{thebibliography}{10}
\bibitem[*]{EHa}
Present Address: Alcatel Space Norway AS, P.O. Box 138, N-3191 Horten,
Norway. 
\bibitem{a1}
H. Munekata, J. C. Maan, L. L. Chang, and L. Esaki, J. Vac. Sci. Technol.
B {\bf 5}, 809 (1987).
\bibitem{a2}
G. Tuttle, H. Kroemer, and J. H. English, J. Appl. Phys. {\bf 65}, 5239
(1989); P. F. Hopkins, A. J. Rimberg, R. M. Westervelt, G. Tuttle, and
H. Kroemer, Appl. Phys. Lett. {\bf 58}, 1428 (1991); Ikai Lo, W. C.
Mitchel, M. O. Manasreh, C. E. Stutz, and K. R. Evans, Appl. Phys. Lett.
{\bf 60}, 751 (1992); J.-P. Cheng, Ikai Lo, and W. C. Mitchel, J. Appl.
Phys. {\bf 76}, 667 (1994); Ikai Lo, W. C. Mitchel, S. Elhamri, R. S.
Newrock, and R. Kaspi, Appl. Phys. Lett. {\bf 65}, 1024 (1994).
\bibitem{a3}
T. P. Smith III, H. Nunekata, L. L. Chang, F. F. Fang, and L. Esaki, Surf.
Sci. {\bf 196}, 687 (1998); Ikai Lo, W. C. Mitchel, and J.-P. Cheng, Phys.
Rev. B {\bf 48}, 9118 (1993).
\bibitem{a4}
J. Kono, B. D. McCombe, J.-P. Cheng, Ikai Lo, W. C. Mitchel, and C. E.
Stutz, Phys. Rev. B {\bf 50}, 12 242 (1994); J.-P. Cheng, J. Kono, B. D.
McCombe, Ikai Lo, W. C. Mitchel, and C. E. Stutz, Phys. Rev. Lett.
{\bf 74}, 450 (1995).
\bibitem{a5}
M. Sweeny and J. Xu, Appl. Phys. Lett. {\bf 54}, 546 (1989).
\bibitem{a6}
J. R. S\"oderstr\"om, D. H. Chow, and T. C. McGill, Appl. Phys. Lett.
{\bf 55}, 1094 (1989).
\bibitem{a7}
L. F. Luo, R. Beresford, and W. I. Wang, Appl. Phys. Lett. {\bf 55},
2023 (1989).
\bibitem{a8}
R. Beresford, L. F. Luo, and W. I. Wang, Appl. Phys. Lett. {\bf 56},
551 (1990).
\bibitem{a9}
R. Beresford, L. F. Luo, K. F. Longenbach, and W. I. Wang, Appl. Phys.
Lett. {\bf 56}, 952 (1990).
\bibitem{a10}
D. A. Collins, E. T. Yu, Y. Rajakarunanayake, J. R. S\"oderstr\"om, D.
Z.-Y. Ting, D. H. Chow, and T. C. McGill, Appl. Phys. Lett. {\bf 57},
683 (1990).
\bibitem{a11}
D. Z.-Y. Ting, D. A. Collins, E. T. Yu, D. H. Chow, and T. C. McGill,
Appl. Phys. Lett. {\bf 57}, 1257 (1990).
\bibitem{a12}
L. Yang, J. F. Chen, and A. Y. Cho, J. Appl. Phys. {\bf 68}, 2997 (1990).
\bibitem{}
D. Z.-Y. Ting, E. T. Yu, and T. C. Mcgill, Phys. Rev. B {\bf 45}, 3583
(1992).
\bibitem{a13}
E. E. Mendez, J. Nocera, and W. I. Wang, Phys. Rev. B {\bf 45}, 3910
(1992).
\bibitem{a14}
M. S. Kiledjian, J. N. Schulman, K. L. Wang, and K. V. Rousseau, Phys.
Rev. B {\bf 46}, 16 012 (1992).
\bibitem{a15}
Maria A. Davidovich, E. V. Anda, C. Tejedor, and G. Platero, Phys. Rev.
B {\bf 47}, 4475 (1993).
\bibitem{a16}
E. E. Mendez, H. Ohno, L. Esaki, and W. I. Wang, Phys. Rev. B {\bf 43},
5196 (1991).
\bibitem{a17}
R. R. Marquardt, D. A. Collins, Y. X. Liu, D. Z.-Y. Ting, and T. C.
McGill, Phys. Rev. B {\bf 53}, 13 624 (1996).
\bibitem{a18}
Y. X. Liu, R. R. Marquardt, D. Z.-Y. Ting, and T. C. McGill, Phys. Rev.
B {\bf 55}, 7073 (1997).
\bibitem{a19}
A. Zakharova, J. Phys.: Condens. Matter {\bf 11}, 4675 (1999).
\bibitem{YangRQ1991}
R. Q. Yang and J. M. Xu, Appl. Phys. Lett. {\bf 59}, 181 (1991).
\bibitem{OhnoH1992}
H. Ohno, L. Esaki, and E. E. Mendez, Appl. Phys. Lett. {\bf 60}, 3153
(1992).
\bibitem{ChangYC1985}
Yia-Chung Chang and J. N. Schulman, Phys. Rev. B {\bf 31}, 2069 (1985).
\bibitem{a20}
Ikai Lo, Jih-Chen Chiang, Shiow-Fon Tsay, W. C. Mitchel, M. Ahoujja, R.
Kaspi, S. Elhamri, and r. S. Newrock, Phys. Rev. B {\bf 55}, 13 677
(1997).
\bibitem{a21}
L.-W. Wang, S.-H. Wei, T. Mattila, A. Zunger, I. Vurgaftman, and J. R.
Meyer, Phys. Rev. B {\bf 60}, 5590 (1999).
\bibitem{NavehY1995}
Y. Naveh and B. Laikhtman, Appl. Phys. Lett. {\bf 66}, 1980 (1995).
\bibitem{ChangYC1988}
Y.-C. Chang, Phys. Rev. B {\bf 37}, 8215 (1988).
\bibitem{ChangYC1990}
Y.-C. Chang, J. Appl. Phys. {\bf 68}, 4233 (1990).
\bibitem{YangRQ1994}
R. Q. Yang, J. M. Xu, and M. Sweeney, Phys. Rev. B {\bf 50}, 7474 (1994).
\bibitem{LuoJ1990}
J. Luo and P. J. Stiles, Phys. Rev. B {\bf 41}, 7685 (1990).
\bibitem{HeidaJP1998}
J. P. Heida, B. J. van Wees, J. J. Kuipers, and T. M. Klapwijk, Phys.
Rev. B {\bf 57}, 11911 (1998).
\bibitem{a23}
R. M. Biefeld, A. A. Allerman, S. R. Kurtz, Mater. Sci. Eng., B {\bf 51},
1 (1998).
\bibitem{a24}
J. R. Meyer, L. J. Olafsen, E. H. Aifer, W. W. Bewley, C. L. felix, I.
Vurgaftman, M. J. Yang, L. Goldberg, D. Zhang, C. H. Lin, S. S. Pei, and
D. H. Chow, IEE Proc.: Optoelectron. {\bf 145}, 275 (1998).
\end{thebibliography}
\end{document}